# A PERFORMANCE EVALUATION OF MOBILE WEB SERVICES SECURITY

Satish Narayana Srirama[1], Matthias Jarke[1,2], Wolfgang Prinz[1,2]
[1]*RWTH Aachen, Informatik V, Ahornstr.55, 52056 Aachen, Germany*
[2]*Fraunhofer FIT, Schloss Birlinghoven, 53754 Sankt Augustin, Germany*
*srirama@cs.rwth-aachen.de, jarke@ cs.rwth-aachen.de, wolfgang.prinz@fit.fraunhofer.de*

Abstract: It is now feasible to host basic web services on a smart phone due to the advances in wireless devices and mobile communication technologies. The market capture of mobile web services also has increased significantly, in the past years. While the applications are quite welcoming, the ability to provide secure and reliable communication in the vulnerable and volatile mobile ad-hoc topologies is vastly becoming necessary. Even though a lot of standardized security specifications like WS-Security, SAML exist for web services in the wired networks, not much has been analyzed and standardized in the wireless environments. In this paper we give our analysis of adapting some of the security standards, especially WS-Security to the cellular domain, with performance statistics. The performance latencies are obtained and analyzed while observing the performance and quality of service of our Mobile Host.

## 1 INTRODUCTION

From the view-point of information systems engineering, the Internet has lead the evolution from static content to Web Services. Web Services are software components that can be accessed over the Internet using well established web mechanisms, XML-based open standards and transport protocols such as SOAP (W3C, 2003) and HTTP. Public interfaces of Web Services are defined and described using Web Service Description Language (WSDL) (Christensen, 2001), regardless of their platforms, implementation details. Web Services have wide range of applications and primarily used for integration of different organizations. The biggest advantage of Web Services lies in their simplicity in expression, communication and servicing. The componentized architecture of Web Services also makes them reusable, thereby reducing the development time and costs. (Booth, 2004)

Simultaneously, the high-end mobile phones and PDAs are becoming pervasive and are being used in wide range of applications like location based services, banking services, ubiquitous computing etc. The higher data transmission rates achieved in wireless domains with 3G (3GPP, 2006) and 4G (Thomas, 1999) technologies also boosted this growth in the cellular market. The situation brings out a large scope and demand for software applications for such high-end mobile devices.

To meet this demand of the cellular domain and to reap the benefits of the fast growing web services domain and standards, the scope of the mobile terminals as both web services clients and providers is being observed. While mobile web service clients are common these days, we have studied the scope of mobile web service provisioning. The details of our Mobile Host and its performance analysis are available at (Srirama, 2006a).

While service delivery and management from Mobile Host is technically feasible, the ability to provide secure and reliable communication in the vulnerable and volatile mobile ad-hoc topologies vastly becomes necessary. Moreover with the easily readable mobile web services, the complexity to realize security increases further. For the traditional wired networks and web services, a lot of standardized security specifications, protocols and implementations like WS-Security (Lawrence, 2004), SAML (Mishra, 2005) etc., exist, but not much has been explored and standardized in wireless environments. Some of the reasons for this poor state might be the lack of active commercial data applications due to the limited resource capabilities of the mobile terminals.

Our study contributes to this work and tries to bridge this gap, with main focus at realizing some of the existing security standards in the mobile web services domain. In this study, we have analyzed the adaptability of WS-Security in the mobile web services domain. Mainly we observed the latency caused to performance of the Mobile Host, with the introduction of security headers into the exchanged SOAP messages. Performance penalties of different encryption and signing algorithms were calculated, and the best possible scenario for securing mobile web services communication is suggested.

The rest of the paper is organized as follows:

Section 2 discusses the standards and on going projects for securing mobile web services domain. Section 3 addresses our security analysis setup and test cases. Section 4 summarizes the results and the means of securing mobile web services. Section 5 concludes the paper with future research directions.

# 2 SECURING MOBILE WEB SERVICE COMMUNICATION

Mobile web service messages are exchanged using the SOAP over different transportation protocols like HTTP, UDP, and WAP etc. SOAP by itself does not specify the means of providing the security for the web service communication. Also many legitimate intermediaries might exist in the web service communication making the security context requirement to be from end-to-end. Hence the traditional point-to-point security technologies like the SSL, HTTPS and full encryption provided by the 3G technologies like UMTS communication technology (Umtsworld, 2002) can't be adapted for the mobile web services domain. These methods also affect the transport independency feature of the SOAP messages by restricting the messages to particular transportation protocols.

For securing the wired web services, OASIS has developed different protocols like WS-Security and SAML using and extending the W3C protocols and standards SOAP, XML Encryption (Reagle, 2001), XML Signature (Eastlake, 2002) and WSDL. WS-Security and SAML protocols make use of the composable and extendable nature of SOAP and embed the security information into SOAP headers.

## 2.1 WS-Security

The WS-Security specification from OASIS is the core element in web service security realm. It provides ways to add security headers to SOAP envelopes, attach security tokens and credentials to a message, insert a timestamp, sign the messages, and encrypt the message. The protocol ensures authentication with security tokens. Security tokens in combination with XML Encryption ensure confidentiality while security tokens in combination with XML Digital Signatures ensure integrity, of the SOAP messages.

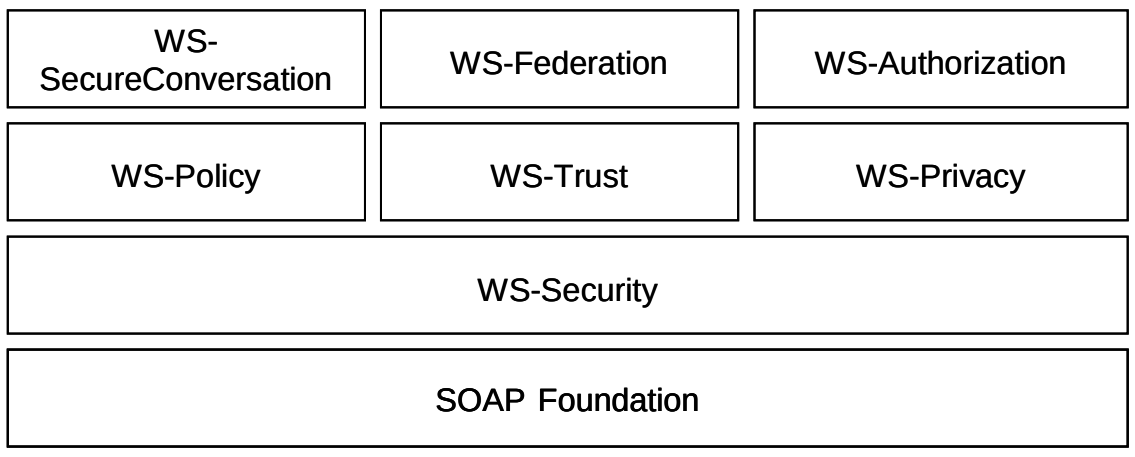

Figure 1: Web Service Security specifications.

Apart from WS-Security, web service security specifications also include WS-Policy which defines the rules for service interaction, WS-Trust which defines trust model for secure exchanges and WS-Privacy which states the maintenance of privacy of information. Built with these set of basic specifications are the specifications, WS-SecureConversation that specifies how to establish and maintain secured session for exchanging data, WS-Federation which defines rules of distributed identity and its maintenance, and WS-Authorization specification which processes the access rights and exchangeable information. The web service security specifications are shown in figure 1. (IBM, 2002)

## 2.2 SAML

Security Assertion Markup Language (SAML) from OASIS primarily provides single sign on (SSO), cross domain interoperability, means of implementing the basic WS-Security standard through assertions, and helps in managing identity control across domains and organizations. SAML builds on top of the web service security specifications and provides a means by which security assertions can be exchanged between different service entity endpoints.

The basic components of interest in SAML are assertions, protocols, bindings and profiles. SAML Assertions carry the authentication information while SAML Request/Response protocols tell how and what assertions can be requested. Bindings define the transportation of SAML protocols over SOAP/HTTP protocol. A SAML profile can be created using the bindings, protocols along with the assertion structure. The SAML Request or SAML Response will reside in SOAP Body.

SAML Request/Response protocol binding over SOAP will provide Assertions in the SOAP Body with information about authentication and authorization. Then SAML Assertions are used along with the WS security element which will reside in SOAP Header. As the SAML Assertions contain key of the holder, it can be used to digitally sign the SOAP Body. At the Receiver end, the signature is verified with the help of the key and the access controls within the Assertion.

## 2.3 LA

Liberty Alliance project (LA, 2006) is the only global body which is working to define and provide technology, knowledge and certifications to build identity into the foundations of mobile and web service communication. It mainly concentrated on federated identity, because of the lack of connectivity between identities for internet applications in the current wireless technology especially in mobile networks.

The basic components of Liberty Alliance are principal, identity provider and service provider. Principal is the requestor who needs to be authenticated. Identity provider is the one which authenticates and asserts the principal's identity. The basic provisions of this project are federation which establishes relationship between any two of the above mentioned components, Single Sign On (SSO) where the authentication provided to principal by the identity provider can be maintained to other components such as service providers, and circle of trust where trust will be established between service providers and identity providers with agreements upon which principals can make transactions and exchange information in a seamless and secure way.

# 3 EVALUATION OF WS-SECURITY FOR SMART PHONES

To secure the communication of our mobile web services provisioning, we have analyzed the adaptability of WS-Security in the mobile web services domain (Srirama, 2006b). The WS-Security adds many performance overheads to the mobile web service invocation cycle. Mainly, extra CPU capabilities are required to process the WS-Security related header elements. The transportation delays also increase significantly as the SOAP message size increases with the added security headers.

During our performance analysis of the Mobile Host, we have observed that the transmission delays sum up to 90% of the total mobile web service invocation cycle times, in GPRS (GSMWorld, 2006) environment. The best solution to cope with this problem would be to increase the transmission capabilities of the wireless networks. With the introduction of 3G technologies like UMTS which promises a data transmission rate of approximately 2 Mbps and the 4G announcement (4GPress, 2005) of achieving the 2.5 Gbps should make the transmission problem void. When such networks are adapted, the increase in the size of the message with security headers is not the major concern.

To evaluate the WS-Security, we observed the latency caused to the performance of the Mobile Host, with the introduction of security headers. The performance penalties of different encryption and signing algorithms were calculated at the smart phone, and the best possible scenario for securing mobile web services communication is observed.

## 3.1 Test setup

The test setup used for WS-Security evaluation is shown in figure 2. The Mobile Host was developed and deployed on a smart phone. The mobile web service client invokes different web services deployed with the Mobile Host. The Mobile Host processes the service request and sends the response back to the client. The performance of the Mobile Host and the network latency were observed while processing the client request.

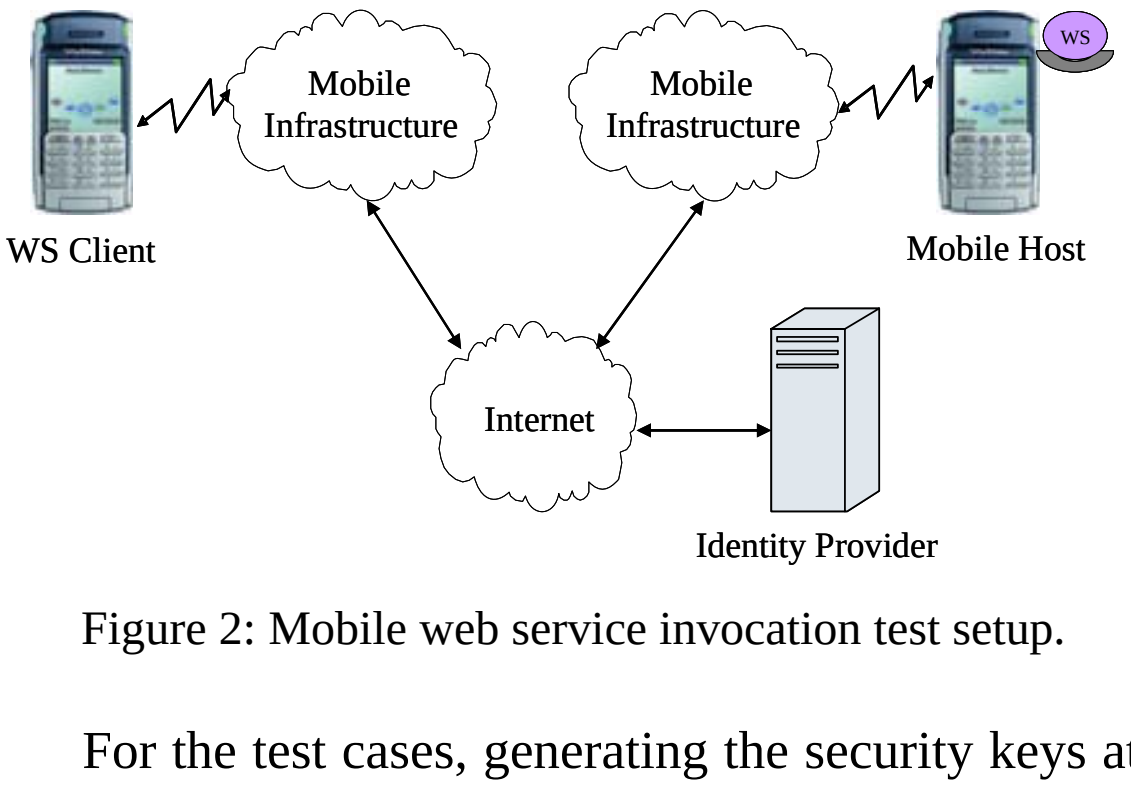

Figure 2: Mobile web service invocation test setup.

For the test cases, generating the security keys at the smart phones was observed to be beyond their low resource capabilities. It was observed that the time taken for generating keys, especially PKI based asymmetric keys is very high. The times were in the order of 1-3 minutes and were highly unpredictable. Hence a standalone end-point security provider (Identity Provider) was used to provide the keys and security information for the secured mobile web service communication. The keys are obtained by connecting to the identity provider across the mobile operator proprietary network and the Internet. The identity provider also helps in achieving SSO. The security assertions and keys were exchanged with the identity provider according to LA standards.

For the analysis, two SonyEricsson P910i smart phones were used as web service requestor and Mobile Host. The smart phones had an internal memory of 64 Kb and ARM9 processor clocked at 156MHz. The phones were connected to the Internet using a GPRS connection. The Mobile Host and client were developed using J2ME MIDP2.0 (JSR118, 2002) with CLDC1.0 (JSR139, 2002) configuration. For cryptographic algorithms and digital signers, java based light weight cryptographic API from Bouncy Castle crypto package (BouncyCastle, 2006) was used. kSOAP2 (KSOAP2, 2006) was modified and adapted according to WS-Security standard and utilized to create the request/response web service messages.

## 3.2 Test cases

Many basic services like the picture service, location data provisioning service, dir service and etc., were provided by the Mobile Host. For the test case analysis of security, addressed in this paper, we considered the location data provisioning service. The service provides the location details of the Mobile Host as GPS information. The smart phone connects to a socket based GPS device via Bluetooth for fetching the GPS co-ordinates. To observe the performance penalties for different message sizes, the request contained the <responseSize> element, which specifies the response size of the message being expected in Kb. The response was also added with an extra element <bodyPadding> to fill the remaining size of the response. The typical request and response of the service are shown in figure 3.

```
<soapenv:Envelope xmlns:soapenv="..."
xmlns:xsd="..." xmlns:xsi="...">
   <soapenv:Body>
     <GPSProvider soapenv:encodingStyle
                    ="...">
      <responseSize xsi:type="xsd:int">
          1 </responseSize>
     </GPSProvider>
    </soapenv:Body>
</soapenv:Envelope>

<SOAP-ENV:Envelope ...>
 <SOAP-ENV:Body ...>
   <GPSProvider xmlns="ssn:SSNServer"
        id="o0" SOAP-ENC:root="1">
    <result>
     <Longtitude xsi:type="xsd:int">
        606428</Longtitude>
     <Latitude ...>5079068</Latitude>
     <Altitude ...>22</Altitude>
     <Speed ...>444</Speed>
      <Status ...>1</Status>
     <Comment xsi:type="xsd:string">
        </Comment>
    </result>
    <Request-ID ...>2</Request-ID>
    <bodyPadding xsi:type="xsd:string">
        ...</bodyPadding>
   </GPSProvider>
 </SOAP-ENV:Body>
</SOAP-ENV:Envelope>
```

Figure 3: Request and response messages of the location data provisioning service, respectively.

To achieve confidentiality, the SOAP messages were ciphered with symmetric encryption algorithms and the generated symmetric keys were exchanged by means of asymmetric public key infrastructure (PKI) based methods. The messages were tested against various symmetric encryption algorithms, along with the WS-Security mandatory algorithms, TRIPLEDES, AES-128, AES-192 and AES-256. (RSA Labs, 2006) RSA with 1024 and 2048 bit keys was used for key exchange. Upon successful analysis of confidentiality, we tried to ensure data integrity of the messages. The messages were digitally signed and were evaluated against two signature algorithms, DSAwithSHA1 (DSS) and RSAwithSHA1 with 1024 and 2048 bit keys. The effect of signing on top of encryption was also studied.

To summarize, the experiments leave us with four test cases for the analysis of WS-Security for smart phones.

Unsecured MWS communication
Encrypted MWS communication
Signed MWS communication
Encrypted + Signed MWS communication

Each of these test cases was observed with different message sizes. The sizes of the response messages ranged from 1-10Kb. All the experiments were repeated at least 5 times and the mean of the values were observed for drawing conclusions, to have statistically valid results.

# 4 ANALYSIS AND THE RESULTS

A typical web service message after applying the WS-Security is shown in figure 4. The SOAP message body can be completely encrypted or only parts of the message can be encrypted. The ciphered data is stored in the body of the updated message. The security information like encryption algorithms used, keys, digests, signing information is maintained in the SOAP header. The message shown below is the snapshot of a message encrypted with AES, and the key exchanged with RSA V 1.5. The message was later signed with RSAwithSHA1.

```
<v:Envelope ...>
 <v:Header>
  <Security>
   <n1:EncryptedKey ...>
    <EncryptedMethod
           Algorithm="...#rsa-1_5" />
    <CipherData> ... </CipherData>
    <ReferenceList>
     <DataReference URI="#4412525"/>
    </ReferenceList>
   </n1:EncryptedKey>
   <n2:Signature ...>
    <SignedInfo> <SignatureMethod
           Algorithm="...#rsa-sha1" />
     <Reference>   <DigestMethod
          Algorithm="...#sha1" />
      <DigestValue>...</DigestValue>
     </Reference>
    </SignedInfo>
    <SignatureValue>
       ...</SignatureValue>
    <KeyInfo> <KeyValue> <RSAKeyValue>
       <Modulus>...</Modulus>
       <Exponent>AQAB</Exponent>
      </RSAKeyValue> </KeyValue>
    </KeyInfo>
   </n2:Signature>
  </Security>
 </v:Header>
 <v:Body>
  <n0:EncryptedData Id="223940028" ...>
   <EncryptionMethod
        Algorithm="...#AESEngine" />
   <CipherData>
    <CipherValue>YeF7...</CipherValue>
   </CipherData>
  </n0:EncryptedData>
 </v:Body>
</v:Envelope>
```

Figure 4: A typical SOAP message with WS-Security.

The example shown in figure 4 also hints the increase in size of the message with the added security header information. With our analysis we have observed that there is a linear increase in the size of the message with the security incorporation. Table 1 shows the variations in mobile web service message size with the applied security. The latency in the encrypted message size for a typical 5 Kb message is approximately 50%.

Table 1: Message size variations (in bytes) with security.

| Original message size | 1024 | 2048 | 5120 | 10240 |
|---|---|---|---|---|
| Message size with Signature | 1726 | 2750 | 5822 | 10942 |
| Encrypted message size | 1804 | 3168 | 7264 | 14092 |
| Secured message size | 2611 | 3975 | 8071 | 14899 |

## 4.1 Encrypted mobile web service communication

To analyze the effects of XML Encryption on the mobile web service invocation cycle, the messages were encrypted with IDEA with 128 and 256 bit keys, DES with 64 and 192 bit keys and AES with 128, 192 and 256 bit keys. The keys were exchanged using RSA with key sizes 1024 and 2048 bits. Figure 5 summaries the results of our encryption analysis and shows the comparison of latency for different encryption algorithms with keys exchanged using RSA 1024. To get the exact performance penalties and to exclude the transmission delays, the invocation cycle was observed on one smart phone.

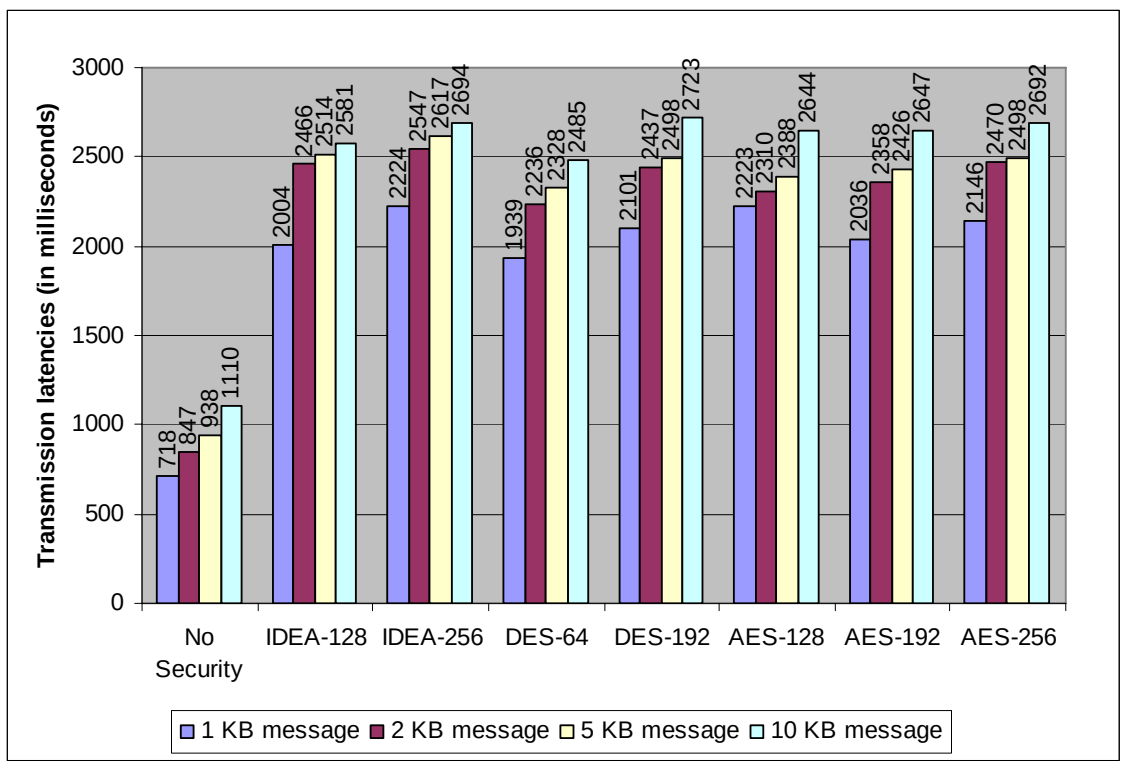

Figure 5: Performance latencies with various symmetric key encryption algorithms and exchanging keys with RSA 1024.

The results suggest that AES 192 encryption turns out be the best symmetric key encryption method. But the difference in latencies with AES

192 and AES 256 are not so significant. So the best means of encrypting the message would be to use AES 256 bit key and to exchange the message with RSA 1024 bit key, both in terms of provided security and performance penalty. Still the increased latency with this best scenario is approximately 3 times the latency without any security. The extra delays mainly constitute the times taken for encryption of request at the client, the decryption of request at the Mobile Host, the encryption of response at the Mobile Host and the decryption of response at the client. RSA 2048 key exchange was beyond the resource capabilities of smart phones.

The results clearly reveal that the processing capabilities of today's smart phones are not yet sufficient for providing proper message level security for mobile web services. Yet the performance penalties are well with in limits, ~= few seconds, and the approach can be used in providing proper security for mobile web services.

## 4.2 Signed mobile web service communication

To analyze the effects of XML digital signatures on the mobile web service invocation cycle, the messages were signed with two signature algorithms, DSAwithSHA1 (DSS) and RSAwithSHA1 with 1024 and 2048 bit key sizes. The latencies are shown in figure 6.

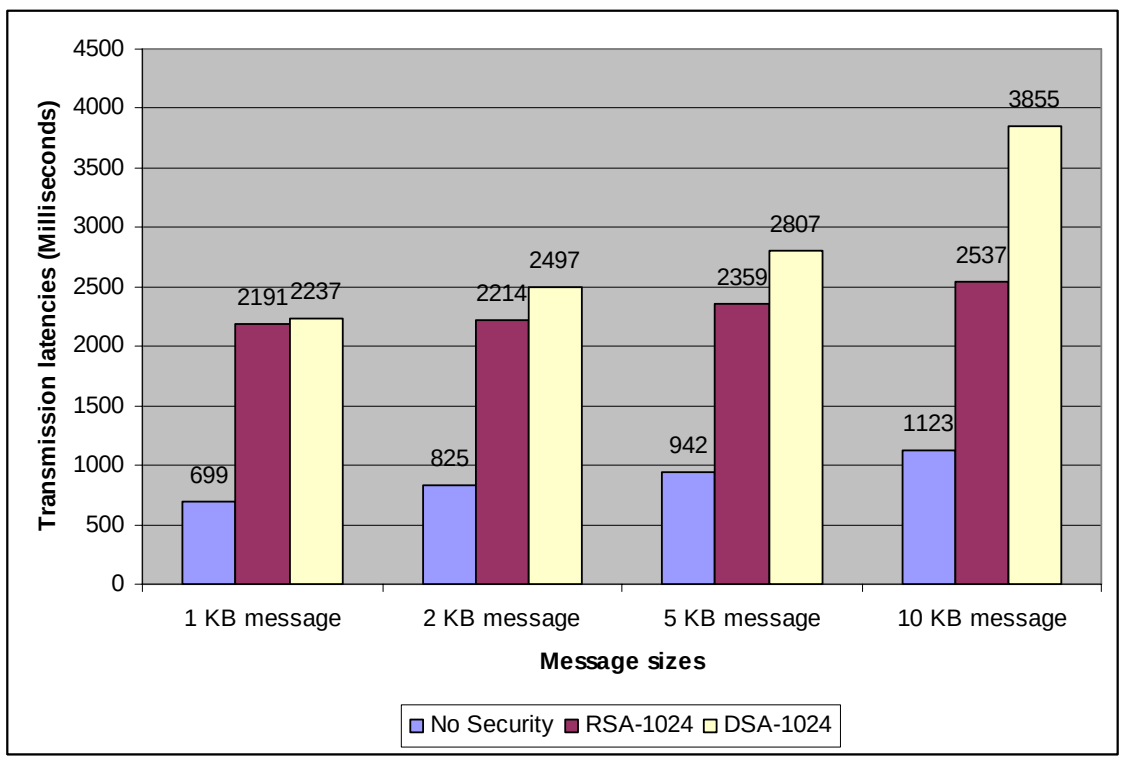

Figure 6: Performance latencies with signing the web service messages.

The results suggest that the best way to sign the mobile web service message would be using RSA V1.5 with 1024 bit key. RSA algorithm is preferred ahead of the DSA, considering performance latencies. With a key size greater than 1024, signing is also beyond the resource capabilities of the smart phone, similar to asymmetric key exchange, using any of the considered algorithms. It was also observed that the latency by signing is slightly higher than the latency caused by the encryption, especially, when considering DSS signature.

## 4.3 Secured mobile web service communication

In real-time applications, to obtain the maximum security and to achieve confidentiality and integrity, both encryption and signing will be used on the messages. So, after the initial individual analysis of encryption and signing, we have checked the latencies for signing on top of encryption. Figure 7 shows the latencies with the complete security applied on the mobile web service messages. The results are for a mobile web service message of request size 1 KB and response size 2 KB.

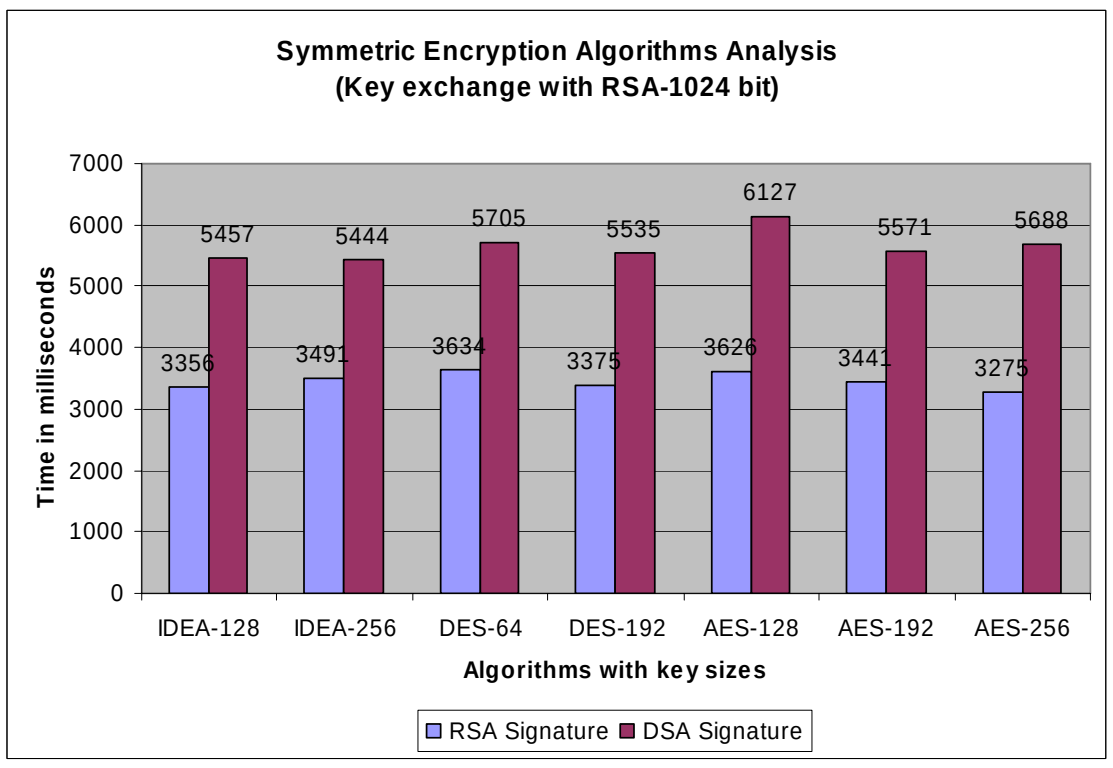

Figure 7: Performance latencies with signing on top of the encryption.

From the test results we can conclude that the best way of securing messages in mobile web service provisioning is to use AES symmetric encryption with 256 bit key, and to exchange the keys with RSA 1024 bit asymmetric key exchange mechanism and signing the messages with RSAwithSHA1. RSA's consideration for signing is not just based on the increased latencies of DSS, but also depended on the asymmetric key exchange mechanism considered. As we were using the RSA for key exchange mechanism, the keys can be reused for signing in the complete secured scenario. This saved us some initializations and thus reduced the latency in the RSA signature test cases.

But there are still high performance penalties when the messages are both encrypted and signed. So we suggest encrypting only the parts of the message, which are critical in terms of security and signing the message. The signing on top of the encryption can also be avoided in specific applications with lower security requirements.

# 5 CONCLUSIONS

The paper summarizes our comprehensive study on analyzing the security for mobile web services provisioning. In this paper we included our analysis of adapting the wired web service security specifications to the cellular world, with performance statistics. The results of our study are welcoming and the mobile web service messages of reasonable size, approximately 2-5kb, can be secured with standard specifications. But based on our till-date realization of security awareness in cellular networks, we conclude that secure web service provisioning in mobile networks is still a great challenge. The mechanisms developed for traditional networks are not always appropriate for the mobile environment and support at hardware like adding an encryption chipset is recommended.

Our future research in this domain includes providing proper end-point security for the Mobile Host with federated identity and appropriate SSO strategy, using SAML and LA standards. We also want to have a detailed performance analysis of the Mobile Host with full security features through real-time applications. We are also looking for alternatives, to reduce the security processing load on the Mobile Host using Enterprise Service Bus (ESB) (Borck, 2005) based mediation framework for maintaining the QoS of the Mobile Host.

The increase in size of the message with the security headers is also quite daunting. We are currently focusing at XML compression and SOAP optimization techniques, to reduce the size of the message, there by improving the scalability of the Mobile Host. The scalability can also be maintained as part of QoS at the mediation framework.

# ACKNOWLEDGEMENTS

The work is supported by German Research Foundation (DFG) as part of the Graduate School ”Software for Mobile Communication Systems” at RWTH Aachen University. The authors also thank R. Levenshteyn and M. Gerdes of Ericsson Research and K. Pendyala for their help and support.